\documentstyle[12pt]{article}
\title{Isospin Mixing and Model Dependence$^*$}
\author{Euy Soo Na and Barry R. Holstein\\[5mm]
Department of Physics and Astronomy\\
University of Massachusetts\\
Amherst, MA 01003}

\begin{document}
\begin{titlepage}
\maketitle

\begin{abstract}
We show that recent calculations of $\Delta I={3\over 2}$ effects in
nonleptonic hyperon decay induced by $m_d-m_u\neq 0$ are subject to
significant model dependence.
\end{abstract}
\vfill
$^*$ Research supported in part by the National Science Foundation.
\end{titlepage}

\section{Introduction}

The isospin breaking caused by the $u,d$ quark mass difference is
well-known and significant.  Indeed the fact the $m_n>m_p$ and the
stability of the proton are a result of this non-degeneracy.  Another
consequence is that mass and isospin eigenstates are not the
same---{\it e.g.} the physical $\Lambda^0$ and $\pi^0$ are admixtures of
the pure $I=0,1$ states $\Lambda_8,\Sigma_3$ and $\pi_8,\pi_3$
respectively.  Since such impurities are small---${\cal
O}(10^{-2})$---we may write\cite{1}
\begin{eqnarray}
\Lambda^0&\approx& \Lambda_8+\theta_b\Sigma_3\nonumber\\
\pi^0&\approx&\pi_3+\theta_m\pi_8
\end{eqnarray}
where the mixing angle is given in terms of quark mass differences as
\begin{equation}
\theta_m=-\theta_b={\sqrt{3}\over 4}{m_d-m_u\over m_s-\hat{m}}
\end{equation}
where $\hat{m}={1\over 2}(m_u+m_d)$.  The size of the quark mass
difference is not completely pinned down, but recent work involving
mesonic mass differences and $\eta\rightarrow 3\pi$ has indicated a
value\cite{2}
\begin{equation}
{m_d-m_u\over m_s-\hat{m}}\approx 0.036
\end{equation}
which corresponds to a mixing angle 
\begin{equation}
\theta_m=-\theta_b\approx 0.016\label{eq:a}
\end{equation}
Perhaps the theoretically cleanest indication of this mixing phenomenon
occurs in the semileptonic $K_{\ell 3}$ decays wherein the ratio of
reduced matrix elements for the decays
\begin{equation}
K^+\rightarrow\pi^0e^+\nu_e\quad{\rm
and}\quad K_L^0\rightarrow\pi^-e^+\nu_e\nonumber
\end{equation}
are found experimentally to be in the ratio \cite{3}
\begin{equation}
\left({f_+^{K^+\pi^0}(0)\over f_+^{K_L^0\pi^-}(0)}\right)^{\rm
exp}=1.029\pm 0.010
\end{equation}
Comparison with the theoretical estimate which arises from mixing 
\begin{equation}
\left({f_+^{K^+\pi^0}(0)\over f_+^{K_L^0\pi^-}(0)}\right)^{\rm theory}
=1+\sqrt{3}\theta_m
\end{equation}
yields a value 
\begin{equation}
\theta_m=0.017\pm 0.005
\end{equation}
quite consistent with Eq. \ref{eq:a} and bears clear witness to the
fact that $\pi^+\pi^0$ are {\it not} exact isotopic partners.

A particularly interesting and important consequence of this mixing
occurs in the arena of nonleptonic weak decays, wherein the
enhancement of $\Delta I={1\over 2}$ transitions by a factor of twenty or
so over their $\Delta I={3\over 2}$ counterparts has long been an item
of study.\cite{4}  The reason that particle mixing effects are particularly 
important in this venue is clear---a 
$\Delta I={1\over 2}$ transition coupled with mixing of
the order of several percent is of the {\it same order} as bona fide
$\Delta I= {3\over 2}$ amplitudes.  Such mixing contributions must
then be subtracted from experimental $\Delta I={1\over 2}$ rule
violating amplitudes before confrontation with theoretical $\Delta
I={3\over 2}$ calculations is made, and such corrections are generally
{\it significant}.  In the case of $K\rightarrow 2\pi$, for example,
we have
\begin{eqnarray}
A(K^+\rightarrow\pi^+\pi^0)&\simeq&\theta_mA(K^+\rightarrow\pi^+\pi_8)\nonumber\\
A(K^0\rightarrow\pi^0\pi^0)&\simeq&A(K^0\rightarrow\pi_3\pi_3)
+2\theta_mA(K^0\rightarrow\pi_3\pi_8)
\end{eqnarray}
The lowest order effective chiral Lagrangian describing this process
is 
\begin{equation}
{\cal L}_w=c_1{\rm tr}\left(\lambda_6D_\mu UD^\mu U^\dagger\right) 
\end{equation}
where 
\begin{equation}
U=\exp\left({i\over F_\pi}\sum_j\lambda_j\phi_j\right)
\end{equation}
is the usual chiral structure, with $F_\pi=92.4$ MeV being the pion
decay constant.\cite{5}  Then we find 
\begin{equation}
A(K^+\rightarrow\pi^+\pi_8)=-\sqrt{2}A(K^0\rightarrow\pi_3\pi_8)
=\sqrt{2\over 3}A(K^0\rightarrow \pi_3\pi_3)
\end{equation}
If we then {\it define} the empirical $\Delta I={3\over 2}$ amplitude
via
\begin{equation}
{}^{3\over 2}d_K^{\rm exp}={3A(K^+\rightarrow\pi^+\pi^0)\over 
2A(K^0\rightarrow\pi^0\pi^0)+A(K^0\rightarrow\pi^+\pi^-)}\approx 0.069
\end{equation}
then the mixing contribution to ${}^{3\over 2}d_K$ is found to be
\begin{equation}
{}^{3\over 2}d_K^{\rm mix}\simeq\sqrt{2\over 3}\theta_m\simeq 0.013
\end{equation}
leaving the isospin ``pure'' piece
\begin{equation}
{}^{3\over 2}d_K^{\rm pure}={}^{3\over 2}d_K^{\rm exp}-{}^{3\over
2}d_K^{\rm mix}\approx 0.056
\end{equation}
This analysis is fairly straightforward and is essentially model-independent,
depending only on the underlying chiral symmetry of QCD.  On the other
hand, things are not so simple in the corresponding hyperon decay
analysis, to which we now turn.

\section{Nonleptonic Hyperon Decay}

In the case of nonleptonic hyperon decay, things are more complex.
Indeed there exist both S-wave (parity-violating) and P-wave
(parity-conserving) amplitudes $A$ and $B$ respectively defined via
\begin{equation}
{\rm Amp}(P\rightarrow P'\pi)=\bar{u}_{P'}(A+B\gamma_5)u_P
\end{equation}
Also, there exist {\it seven} different channels with $\Delta I={1\over 2}$
and $\Delta I={3\over 2}$ components in each.  We define $\Delta
I={3\over 2}$ parameters via
\begin{eqnarray}
c_\Lambda^{3\over 2}&=&-\sqrt{1\over 3}\left[{\rm Amp}(\Lambda^0\rightarrow
p\pi^-)+\sqrt{2}{\rm Amp}(\Lambda^0\rightarrow n\pi^0)\right]\nonumber\\
c_\Sigma^{3\over 2}&=&{\rm Amp}(\Sigma^+\rightarrow n\pi^+)-{\rm
Amp}(\Sigma^-\rightarrow n\pi^-)-\sqrt2{{\rm Amp}(\Sigma^+\rightarrow
p\pi^0)}\nonumber\\
c_\Xi^{3\over 2}&=&-{2\over 3}\left[{\rm
Amp}(\Xi^-\rightarrow\Lambda^0\pi^-)
+\sqrt{2}{\rm Amp}(\Xi^0\rightarrow\Lambda^0\pi^0)\right]
\end{eqnarray}
for $A,B$ amplitudes respectively.  The experimental values for each
quantity are given in Table 1,\cite{4} where all quoted numbers are in units
of $10^{-7}$.  Since corresponding $\Delta I={1\over 2}$ quantities
are of order 5-15 ($\times 10^{-7}$) the $\Delta I={3\over 2}$
suppression is clear.

In order to estimate the mixing contributions to these parameters,
one needs a realistic model for nonleptonic hyperon decay and this is
where the problem lies.  Indeed in the standard picture
S-wave amplitudes are given by the PCAC-commutator contributions\cite{6}
\begin{eqnarray}
<\pi^aP'|{\cal H}_w^{PV}|P>&=&-{i\over F_\pi}
<P'|[F^5_a,{\cal H}_w^{PV}]|P>\nonumber\\
&=&-{i\over F_\pi}<P'|[F_a,{\cal H}_w^{PC}]|P>
\end{eqnarray}
while P-waves are represented by baryon pole terms
\begin{eqnarray}
<\pi^aP'|{\cal H}_w^{PC}|P>&=&\sum_{P''}<\pi^aP'|P''>{i\over
m_P-m_{P''}}<P''|{\cal H}_w^{PC}|P>\nonumber\\
&+&\sum_{P''}<P'|{\cal H}_w^{PC}|P''>{i\over m_{P'}-m_{P''}}<\pi^aP''|P>
\end{eqnarray}
The weak parity-conserving baryon-baryon amplitudes are characterized via
SU(3) $F,D$ couplings as
\begin{equation}
<P_j|{\cal H}_w^{PC}|P_i>=\bar{u}_j(-if_{6ij}F+d_{6ij}D)u_i
\label{eq:aa}
\end{equation}
The strong mesonic coupligns are represented in terms of the
generalized Goldberger-Treiman relation as\cite{7}
\begin{equation}
g_A^{ijk}={2F_\pi\over m_j+m_k}g^{ijk}
\end{equation}
with the pseudoscalar couplings $g^{ijk}$ given in terms of SU(3)
$f,d$ couplings as
\begin{equation}
g^{ijk}=-2(-if_{ijk}f+d_{ijk}d)g
\end{equation}
with $g^2/4\pi\approx 13$.  Then, for example, we have
\begin{eqnarray}
A(\Sigma^+\rightarrow p\pi^0)&=&{1\over F_\pi}(D-F)\nonumber\\
A(\Lambda^0\rightarrow p\pi^-)&=&{1\over \sqrt{3}F_\pi}(D+3F)\nonumber\\
&etc.&
\end{eqnarray}
for S-wave amplitudes and 
\begin{eqnarray}
B(\Sigma^+\rightarrow p\pi^0)&=&2g(m_N+m_\Sigma)
({(f+d)(F-D)\over 2m_N(m_\Sigma-m_N)}\nonumber\\
&-&{2f(F-D)\over 2m_\Sigma(m_\Sigma-m_N)})\nonumber\\
B(\Lambda^0\rightarrow p\pi^-)&=&{2\over \sqrt{3}}g(m_N+m_\Lambda)
({(f+d)(3F+D)\over 2m_N(m_\Lambda-m_N)}\nonumber\\
&-&{2d(F-D)\over (m_\Sigma+m_\Lambda)(m_\Sigma-m_N)})\nonumber\\
&etc.&
\end{eqnarray}
for P-waves.  In the case of the strong couplings the values\cite{8} 
\begin{equation}
f+d=1, \quad {d\over f}=1.8
\end{equation}
are generally accepted.  However, there is no consensus for the weak
parameters $F,D$.  If one employs the values $D/F=-0.42$ and
$F/2F_\pi=1.13\times 10^{-7}$ 
which provide a good fit to S-wave terms $A$ then a poor fit is given
for P-waves as shown as ``model 1'' in Table 1.  
On the other hand, using $D/F=-0.85$
and $F/2F_\pi=1.83\times 10^{-7}$ yields a good 
P-wave representation but a poor S-wave
fit---{\it cf.} ``model 2'' in Table 1.\cite{8}  
This problem has been known for a long time,
and a definitive and widely accepted solution has yet to be found.  
One intriguing possibility was put forth by LeYaouanc et al., who
point out that a reasonable fit to {\it both} S- and P-wave amplitudes 
can be provided ({\it cf.} ``model 3'' in {Table 2) by 
appending intermediate state contributions from SU(6)
$70,1^-$ states to usual S-wave commutator terms.\cite{10}  Such contributions,
of course, vanish in the soft pion limit if SU(3) invariance obtains, 
but in the real world such contributions can be sizable and when
estimated using a simple constituent quark model seem to be able to
provide a satisfactory resolution to the S/P dilemma.  Of course, this
suggestion is not unique and other possibilities have been proposed.
However, our purpose in this note is not to provide a solution to the
problem of hyperon decay but rather to study the model dependence of
the mixing estimates. 
 
\begin{table}
\begin{center}
\begin{tabular}{c|cccc|cccc}
      &  & S-waves & & & P-waves & \\
      & exp & model 1 & model 2 &model 3& exp & model 1 & model
2&model 3\\
\hline
$\Lambda^0_-$& 3.25 & 3.36 &4.55& 3.21 & 22.1& 30.6& 25.9&25.9\\
$\Sigma^+_0$&-3.27 &-3.20 &-6.78 &-3.44& 26.6& 15.4& 32.6&32.6 \\
$\Sigma^-_-$&4.27&4.53&9.59 &4.87& -1.44 & -8.7&-1.1&-1.1\\
$\Xi^-_-$&-4.51&-4.45&-8.15 &-5.08& 16.6 & -5.9&17.6&17.6\\
\hline
\end{tabular}
\end{center}
\caption{Shown are values for the S-wave hyperon decay amplitudes $A$
for various channels as obtained experimentally and in models.  All
numbers are to be multiplied by $10^{-7}$.  Models 1,2,3 are described
in the text.}
\end{table}

In these various pictures of hyperon decay one can easily calculate
the size of the mixing contributions to the experimental $\Delta
I={1\over 2}$ rule violating parameters $c_i^{3\over 2}$.  
In order to accomplish this program one
requires various unphysical weak decay amplitudes, but these are
straightforwardly calculated in the various models, yielding the
results 
\begin{eqnarray}
A(\Sigma^+\rightarrow p\pi_8)&=&-{\sqrt{3}\over F_\pi}(D-F)
+{1\over F_\pi}(m_\Sigma-m_N)30C\nonumber\\
A(\Lambda^0\rightarrow n\pi_8)&=&{1\over
\sqrt{2}F_\pi}(D+3F)+{1\over F_\pi}(m_\Lambda-m_N)3\sqrt{6}C\nonumber\\
A(\Sigma^0\rightarrow p\pi^-)&=&{1\over F_\pi}(-D+F)
-{1\over F_\pi}(m_\Sigma-m_N)18\sqrt{3}C\nonumber\\
A(\Xi^0\rightarrow\Lambda^0\pi_8)&=&-{1\over
\sqrt{2}F_\pi}(-D+3F)-{1\over F_\pi}(m_\Xi-m_\Lambda)2\sqrt{6}C
\end{eqnarray}
where $F,D$ are the weak decay parameters defined in Eq. \ref{eq:aa} and 
\begin{equation}
C={1\over
4\sqrt{3}}G\cos\theta_C\sin\theta_C
{<\psi^s|\delta(r_1-r_2)|\psi^s>\over m^2R^2\omega}
\end{equation}
is a parameter defined by Le Yaouanc et al. which arises from the $1^-$
intermediate state contributions.  From the S-wave fit given in Table 1
one determines $C\simeq 3.9\times 10^{-9}$ and can then calculate the
various contributions to $d_i^{3\over 2}$, yielding the results shown
in Table 2.
\begin{table}
\begin{center}
\begin{tabular}{c|c|c|c|c}
 & exp&model 1&model 2&model 3\\
\hline \\
$A_\Lambda^{3\over 2}$&0.059&-0.005 &-0.113 &-0.048\\
$A_\Xi^{3\over 2}$&-0.227&-0.051&-0.081 &-0.098 \\
$A_\Sigma^{3\over 2}$&0.485&0.118&0.249 &0.317 \\
\hline
$B_\Lambda^{3\over 2}$&0.141&0.500& 0.546&0.546 \\
$B_\Xi^{3\over 2}$&0.530&0.504&0.791 &0.791 \\
$B_\Sigma^{3\over 2}$&6.022&-0.256&-0.542 &-0.542 
\end{tabular}
\end{center}
\caption{Shown are the predicted values of $\Delta I=3/2$ amplitudes
for both parity conserving and violating sectors of the hyperon decays
compared to their experimental values.  Models 1,2,3 are described in
the text.}
\end{table}

Study of the numbers given in this table reveals the point of our
note---mixing contributions to $\Delta I={3\over 2}$ weak decay amplitudes
are of the same size as the experimental numbers themselves {\it and}
are quite model dependent.  Indeed, Maltman recently calculated the
values given for model 1, obtaining numbers which represent generally
$\sim 25\%$ corrections for S-waves and $\sim 100\%$ corrections for
P-waves.\cite{11}  We see, however, that results can be very different 
for models which are equally capable or describing the hyperon decay
data.  For instance, in the successful model of LaYouanc et al. the
corrections in both S- and P-wave channels are found to be $\sim
100\%$, while we see from comparison of models 1 and
2 that even in the
basic model the results are very sensitive to the values for the weak F,D
coefficients which are chosen.  We do not claim here then to reliably 
calculate the size of the simulated $\Delta I={3\over 2}$ effect---rather
to merely note the rather significant model dependence of same.  This
result has interesting implications for those attempting to calculate
bona fide $\Delta I={3\over 2}$ effects in nonleptonic decays when
comparison with experiment is attempted, but those are the subject of
another paper.

\end{document}